\documentclass[12pt]{article}
\usepackage{amssymb}
\usepackage{amsfonts}
\usepackage{amsmath}
\usepackage{amsmath,amsthm}
\usepackage{graphicx,epsfig}
\RequirePackage[numbers,sort&compress]{natbib}

\begin{document}

\setcounter{page}{1}

\pagestyle{plain} \vspace{1cm}

\begin{center}
\Large{\bf Energy conditions in $f(T)$ gravity with higher-derivative torsion terms}\\
\small \vspace{1cm}{\bf Tahereh Azizi \footnote{t.azizi@umz.ac.ir}and \bf Miysam Gorjizadeh \footnote{miysamgorjizadehbaee@yahoo.com}}\\
\vspace{0.5cm} Department of Physics, Faculty of Basic
Sciences,\\
University of Mazandaran,\\
P. O. Box 47416-95447, Babolsar, IRAN\\

\end{center} \vspace{1.5cm}

\begin{abstract}

We study the energy conditions in the framework of the modified gravity with
higher-derivative torsional terms in the action. We discuss the viability of the model by studying the energy conditions in terms
 of the cosmographical parameters like Hubble, deceleration, jerk, snap and lerk parameters.
In particular, We consider two specific models that are proposed in literature and examine
the viability bounds imposed by the weak energy condition.
\end{abstract}

\section{Introduction}
Recent astronomical observations \cite{Riess98,Perlmutter99,Ade15} have shown that the current universe experiences
an accelerated expansion. To explain this unexpected phenomenon, two remarkable approaches
have been suggested. In the first approach, an exotic matter component with negative pressure is considered
in the right hand side of the Einstien equations dubbed as, dark energy in the literature.
There are several candidates for the dark energy proposal such as the cosmological
constant \cite{Peebles03}, canonical scalar field (quintessence) \cite{Li02},  phantom field \cite{Caldwell02},
chaplygin gas \cite{Kamenshchik01} and so on (for a review see \cite{Copeland06}).
The second approach is based on the modifying the left hand side of the field
 equations dubbed as, dark gravity. Some examples of the modified gravity models are
 $f(R)$ theories, string inspired gravity, braneworld gravity, etc ( for review see \cite{Sotiriou10,Nojiri11,Clifton12} and
references therein). One of the interesting modified gravity model is the $f(T)$ gravity which $T$ is
the torsion scalar. This scenario is based on the “teleparallel” equivalent of General Relativity (TEGR)
\cite{Einstein28,Hayashi79} which uses the Weitzenb\"{o}ck connection
that has no curvature but only torsion. Note that the Lagrangian density of the Einstien gravity is
constructed from the curvature defined via the Levi-Civita connection.
In the context of the TEGR, the dynamical object is
a vierbein field $e_i(x^{\mu})$, $i=0,1,2,3$,
which form the orthogonal bases for the tangent
space at each point of spacetime. The metric tensor is obtained
from the dual vierbein as $g_{\mu\nu}(x)=\eta_{ij}e_{\mu}^{i}(x)e_{\nu}^{j}(x)$
where $\eta_{ij}=e_{i}.e_{j}$ is the Minkowski metric and
$e^{\mu}_{i}$ is the component of the vector $e_i$ in a coordinate basis.
 Note that the Greek indices are refer to the coordinates on the
manifold while Latin indices label the tangent space. The Lagrangian density of the teleparallel gravity
 is constructed from the torsion tensor which is defined as
\begin{equation}
T^{\lambda}_{~\mu\nu}=e^{\lambda}_{i}(\partial_{\mu}e_{\nu}^{i}-\partial_{\nu}e_{\mu}^{i})\nonumber.\label{tele1}
\end{equation}
One can write down the torsion scalar $T\equiv S^{~\mu\nu}_{\lambda}T_{~\nu\mu}^{\lambda}$ where
\begin{equation}
S^{~\mu\nu}_{\lambda}\equiv\frac{1}{2}(K^{\mu\nu}_{~\lambda}+\delta_{\lambda}^{\mu}T^{\sigma\nu}_{~\sigma}-
\delta_{\lambda}^{\nu}T^{\sigma\mu}_{~\sigma})\nonumber\label{tele2}
\end{equation}
and $K^{\mu\nu}$ is the contorsion tensor defined as
\begin{equation}
K^{\mu\nu}_{~\lambda}=-\frac{1}{2}(T^{\mu\nu}_{~\lambda}-T^{\nu\mu}_{~\lambda}-T^{~\mu\nu}_{\lambda})\,.\nonumber\label{tele3}
\end{equation}
Using the torsion scalar as the teleparallel Lagrangian leads to the same gravitational equations of the
 general relativity.

 Similar to the $f(R)$  modified gravity, one can modify
 teleparallel gravity by considering an arbitrary function of the torsion scalar
 in the action of the theory, which leads to the $f(T)$ theories of gravity \cite{Ferraro07,Bengochea09,Linder10}.
 It is worth noticing that the field equations of the $f(T)$ gravity are second order differential equations
 and so it is more manageable compared to the $f(R)$ theories whose the field equations are $4$nd order equations.
  Consequently, the  modified TEGR  models have attracted a lot of interest in literature
  (see \cite{Cai16} and references therein ).
 Recently, a further modification of the teleparallel gravity has been proposed with constructing
  a torsional gravitational modifications using higher-derivative terms such as $(\nabla T)^{2}$ and $\Box T$
  terms in the Lagrangian of the theory \cite{Otalora16}.
  In this regard, the dynamical system of this model has been studied
   via performing the phase-space analysis of the cosmological
scenario and consequently, an effective dark energy sector that
  comprises of the novel torsional contributions is obtained.

The aim of this paper is to explore the energy conditions of this modified teleparallel
 gravity by taking into account the further degrees of freedom related to the higher derivative terms.
  Indeed, one procedure for analyzing the viability of the modified gravity models is studying
  the energy conditions to constrain the free parameters of them. In this respect, one can impose the null,
  weak, dominant and strong energy conditions ,which arise from the Raychaudhuri equation for the
   expansion \cite{Dadhich07,Kar07}, to the modified gravity model.
  In the literature, this approach has been extensively studied to evaluate the possible
  ranges of the free parameter of the generalized gravity models. For instants, the energy bound have been explored to
  constrain $f(R)$ theories of gravity \cite{Ber06,Sant07,Ata09,Capozziello15} and some extensions of $f(R)$ gravity \cite{Bertolami09,Garcia10,Wang10,Garcia11,Wang12,Alvarenga13,Haghani13,Odintsov13,Sharif13},
  modified Gauss-Bonnet gravity \cite{Wu10,Garcia11a,Garcia11b,Bani12} and scalar-tensor gravity \cite{Sharif13a,Capozziello14}.
   The energy condition have also been analysed in $f(T)$ gravity \cite{Liu12,Boehmer12,Jamil13}
    and generalized models of $f(T)$ gravity \cite{Kiani14,Zubair15,Jawad15,Sharif16}.

  To study the energy conditions in this modified teleparallel gravity,
  we consider the flat FRW universe model with perfect fluid matter
  and define an effective energy density and pressure
  originates from the higher-derivative torsion terms. Then we discuss the energy conditions in term of the cosmographical parameters such as
  the Hubble, deceleration, jerk,
   snap and lerk parameters. Particularly, we consider two specific models that are proposed in literature and
  using the present-day values of the cosmographical parameters, we analyze the weak energy
  condition to determine the possible constraints on the free parameter of the presented models.

   The paper is organized as follows: In section 2 we review the modified $f(T)$ gravity with
   higher-derivative torsion terms,
    the equations of motion and the resulted modified Friedmann equations
    related to the model. Section 3 is devoted to the energy conditions in this modified teleparallel gravity.
     In section 4, we explore the weak energy condition in two specific models of the scenario
     by using present-day values of the cosmographic quantities.
     Finally, our conclusion will be appeared in section 4.
\section{The field equations}

The action of the modified teleparallel gravity with higher-derivative torsion terms is defined as \cite{Otalora16}
\begin{equation}
S=\frac{1}{2}\int \,{\mathrm{d^{4}}}x |e| F\big(T,(\nabla T)^{2},\Box T\big)+S_{m}(e^{A}_{\rho},\Psi_{m}),\label{action}
\end{equation}
where $e=det(e^{A}_{\mu})=\sqrt{-g}$, $(\nabla T)^{2}=\eta^{AB}e^{\mu}_{A}e^{\nu}_{B}\nabla _{\mu}T\nabla_{\nu}T=
g^{\mu\nu}\nabla _{\mu}T\nabla_{\nu}T$ and $\Box T=\eta^{AB}e^{\mu}_{A}e^{\nu}_{B}\nabla_{\mu}\nabla_{\nu}T=
g^{\mu\nu}\nabla_{\mu}\nabla_{\nu}T$, and for simplicity we have set $\kappa^{2}=8\pi G=1$. The $S_{m}(e^{A}_{\rho},\Psi_{m})$
is the matter action includes the general matter field which can ,in general,
 have an arbitrary coupling to the vierbein field.
If the matter couples to the metric in the standard form, then varying the action (\ref{action})
 with respect to the vierbein yields the generalized field equations as follows\cite{Otalora16}

\begin{eqnarray}
\begin{split}
&\frac{1}{e}\partial_{\mu}(eF_{T}e^{\sigma}_{A}S^{\rho\mu}_{\sigma})-F_{T}e^
{\sigma}_{A}S^{\mu\rho}_{\nu}T^{\nu}_{\mu\sigma}+\frac{1}{4}e^{\rho}_
{A}F-\frac{1}{4e}\partial_{\lambda}\partial_{\mu}\partial_{\nu}\Big(eF_{X_2}
\frac{\partial _{X_2}}{\partial_{\lambda}\partial_{\mu}\partial_{\nu}e^{A}_{\rho}}\Big)
\\&+\frac{1}{4}\sum_{a=1}^{2}\left\{{{F_{X_a}\frac{\partial_{ X_a}}{\partial e^{A}_{\rho}}-
\frac{1}{e}\left[\partial_{\mu}\Big(eF_{X_a}\frac{\partial_{X_a}}{\partial \partial_{\mu}e^{A}
_{\rho}}\Big)-\partial_{\mu}\partial_{\nu}\Big(eF_{X_a}\frac{\partial X_a}
{\partial\partial_{\mu}\partial_{\nu}e^{A}_{\rho}}\Big)\right]}}\right\}
\\&=\frac{1}{2}e^{\sigma}_{A}\mathcal{T}^{(m)\rho}_{\sigma},\label{field1}
\end{split}
\end{eqnarray}
where for simplicity, we have used the notation $X_{1}\equiv(\nabla T)^{2}$ and $X_{2}\equiv \Box T$.
Note that $F_{T}$ and $F_{X_a}$ denote derivative with respect to the
torsion scalar and $X_a$, with $a=1,2$, respectively and
$\mathcal{T}^{(m)}_{\rho\sigma}$ is the matter energy momentum tensor which
is defined as $e^{\sigma}_{A}\mathcal{T}^{(m)\rho}_{\sigma}
\equiv -\frac{1}{e}\frac{\delta S_{m}}{\delta e^{A}_{\rho}}$.
In order to study the cosmological implication of the model, we consider a
spacially flat Friedmann-Robertson-Walker (FRW) universe with metric $ds^{2}=dt^{2}-a^{2}(t)\delta_{ij}dx^{i}dx^{j}$.
This metric arises from the following diagonal vierbein
\begin{equation}
e^{A}_{\mu}=diag\big(1,a(t),a(t),a(t)\big),\label{vierbien-FRW}
\end{equation}
where $a(t)$ is the scale factor. Now we assume that the matter content of
the universe is given by the perfect fluid with the energy
 density $\rho_{m}$ and pressure $p_{m}$. Thus using the field equations  (\ref{field1}),
 we obtain the generalized Friedmann equations as follows
\begin{eqnarray}
\begin{split}
&F_{T}H^{2}+(24H^{2}F_{X_1}+F_{X_2})\left(3H\dot{H}+\ddot{H}\right)H+F_{X_2}\dot{H}^{2}\\
&+\left(3H^{2}-\dot{H}\right)H\dot{F}_{X_2}+
24H^{3}\dot{H}\dot{F}_{X_1}+H^{2}\ddot{F}_{X_2}+\frac{F}{12}=\frac{\rho_{m}}{6},\label{friedman1}
\end{split}
\end{eqnarray}

\begin{eqnarray}
\begin{split}
&F_{T}\dot{H}+H\dot{F}_{T}+24H\left[2H\ddot{H}+3(\dot{H}+H^{2})\dot{H}\right]
\dot{F}_{X_1}+12H\dot{H}\dot{F}_{X_2}+24H^{2}\dot{H}\ddot{F}_{X_1}\\
&+\left(\dot{H}+3H^{2}\right)\ddot{F}_{X_2}+24H^{2}F_{X_1}\dddot{H}+
H\dddot{F}_{X_2}+24F_{X_1}\dot{H}^{2}\left(12H^{2}+\dot{H}\right)\\
&+24HF_{X_1}\left(4\dot{H}+3H^{2}\right)\ddot{H}=-\frac{p_{m}}{2},\label{friedman2}
\end{split}
\end{eqnarray}
where dot denotes the derivative with respect to cosmic time and $H=\dot{a}/a$ is the Hubble parameter. With the
definition of the veirbien (\ref{vierbien-FRW}), the torsion scalar
and the functions $X_{1}$ and $X_{2}$ respectively are given by
\begin{equation}
T=-6H^{2},\label{tor-scalar}
\end{equation}
\begin{equation}
X_{1}=144H^{2}\dot{H}^{2},\label{X_1}
\end{equation}
\begin{equation}
X_{2}=-12\left[\dot{H}\big(\dot{H}+3H^{2}\big)+H\ddot{H}\right].\label{X_2}
\end{equation}
Substituting (\ref{tor-scalar}), (\ref{X_1}) and (\ref{X_2}) in the equations (\ref{friedman1})
and (\ref{friedman2}), we can rewrite the Freidmann equations in the usual form as general relativity
\begin{equation}
3H^{2}=\rho_{de}+\rho_{m},\label{friedman3}
\end{equation}
 \begin{equation}
-2\dot{H}=\rho_{m}+p_{m}+\rho_{de}+p_{de},\label{friedman4}
\end{equation}
where the energy density and pressure of the effective dark energy sector are respectively defined as
\begin{eqnarray}
\begin{split}
&\rho_{de}\equiv-\frac{F}{2}-6H^{2}F_{T}+3H^{2}-6F_{X_2}\dot{H^{2}}-6H(3H^{2}-\dot{H})\dot{F}_{X_2}\\
&-144H^{3}\dot{H}\dot{F}_{X_1}-6H\left(F_{X_2}+24H^{2}F_{X_1}\right)\left(3H\dot{H}+\ddot{H}\right)-6H^{2}\ddot{F}_{X_2},\label{rho}
\end{split}
\end{eqnarray}

\begin{eqnarray}
\begin{split}
&p_{de}\equiv-3H^{2}-2\dot{H}+2\Big[F_{T}\dot{H}+H\dot{F}_{T}+24H\left(2H\ddot{H}+3(\dot{H}+H^{2})\dot{H}\right)\dot{F}_{X_1}\\
&+12H\dot{H}\dot{F}_{X_2}+24H^{2}\dot{H}\ddot{F}_{X_1}+\big(\dot{H}+3H^{2}\big)\ddot{F}_{X_2}+24H^{2}F_{X_1}\dddot{H}+H\dddot{F}_{X_2}\\
&+24F_{X_1}\dot{H^{2}}(12H^{2}+\dot{H})+24HF_{X_1}\big(4\dot{H}+3H^{2}\big)\ddot{H}\Big].\label{p}
\end{split}
\end{eqnarray}
Since we have assumed a minimally coupled matter to the veirbien field,
the standard matter is satisfied in continuity equation , i.e. $\dot{\rho}_{m}+3H(p_{m}+\rho_{m})=0$.
So from the friedmann equations (\ref{friedman3}) and (\ref{friedman4}),
one can easily verified that the dark energy density and pressure satisfy the conservation equation
 \begin{equation}
\dot{\rho}_{de}+3H(p_{de}+\rho_{de})=0.
\end{equation}
In the rest of this paper we analyze the viability of this modified teleparallel gravity scenario by studying
  the energy conditions to constrain the free parameters of the model.

\section{Energy conditions}
The energy conditions are originated from the Raychaudhuri equation together with the requirement
that gravity is attractive for a space-time manifold which is endowed by a metric $g_{\mu\nu}$.
In the case of a congruence of timelike and null geodesics with tangent vector field $u^{\mu}$ and $k^{\mu}$ respectively,
 the Raychaudhuri equation is given
by the temporal variation of expansion for the respective curve as follows
 \cite{Dadhich07,Kar07,Capozziello15}
\begin{equation}
\frac{d\theta}{d\tau}=-\frac{1}{3}\theta^{2}-\sigma_{\mu\nu}\sigma^{\mu\nu}+\omega_{\mu\nu}\omega^{\mu\nu}-R_{\mu\nu}u^{\mu}u^{\nu}\,,
\end{equation}

\begin{equation}
\frac{d\theta}{d\tau}=-\frac{1}{2}\theta^{2}-\sigma_{\mu\nu}\sigma^{\mu\nu}+\omega_{\mu\nu}\omega^{\mu\nu}-R_{\mu\nu}k^{\mu}k^{\nu}\,.
\end{equation}
Here $R_{\mu\nu}$ is the Ricci tensor and $\theta$, $\sigma^{\mu\nu}$ and $\omega_{\mu\nu}$ are, respectively, the expansion scalar,
shear tensor and rotation tensor associated with the congruence of timelike or null geodesics.
Note that the Raychaudhuri equation is a purely geometric equation hence,
it makes no reference to a specific theory of gravitation. Since the shear is a purely spatial tensor
($\sigma_{\mu\nu}\sigma^{\mu\nu}\geq0$), for any hypersurface of orthogonal congruence
($\omega_{\mu\nu}=0$), the conditions for attractive gravity reduces to
 \begin{eqnarray}
 R_{\mu\nu}u^{\mu}u^{\nu}\geq0,~~~
R_{\mu\nu}k^{\mu}k^{\nu}\geq0,\label{energy-cond1}
\end{eqnarray}
where the first condition is refer to the strong energy condition (SEC) and the second condition is named null energy condition (NEC).
From the field equations in general relativity and its modifications, the Ricci tensor is related to the energy-momentum tensor
of the matter contents.
 Thus inequalities (\ref{energy-cond1}) give rise to the respective physical conditions
 on the energy-momentum tensor as
\begin{eqnarray}
\left(\mathcal{T}_{\mu\nu}-\frac{\mathcal{T}}{2}g_{\mu\nu}\right)u^{\mu}u^{\nu}\geq0,\quad
\left(\mathcal{T}_{\mu\nu}-\frac{\mathcal{T}}{2}g_{\mu\nu}\right)k^{\mu}k^{\nu}\geq0,
\end{eqnarray}
 where $\mathcal{T}$ is the trace of the energy-momentum tensor. For perfect fluid
 with energy density $\rho_m$ and pressure $p_m$, the
 SEC and NEC are defined by $\rho_m+3p_m\geq0$ and $\rho_m+p_m\geq0$ respectively,
 while the dominant energy condition (DEC) and weak energy condition (WEC)
 are defined respectively, by $\rho_m\pm p_m\geq0$ and $\rho\geq0$. Note that
 the violation of the NEC leads to the violation of all other
 conditions. Since the Raychaudhuri equation is a purely geometric equation, the concept
 of energy conditions can be extended to the case of modified
theories of gravity with the assumption that the total matter contents of the universe act
like a perfect fluid. Hence, the respective conditions can be defined by replacing
the energy density and pressure with an effective energy density
and effective pressure, respectively as follows
 \begin{eqnarray}
 \begin{split}
&\textbf{NEC}:\quad\rho_{eff}+p_{eff}\geq0,\\
&\textbf{SEC}:\quad\rho_{eff}+p_{eff}\geq0,\quad\rho_{eff}+3p_{eff}\geq0,\\
&\textbf{DEC}:\quad\rho_{eff}\geq0,\quad\rho_{eff}\pm p_{eff}\geq0,\\
&\textbf{WEC}:\quad\rho_{eff}\geq0,\quad\rho_{eff}+p_{eff}\geq0.\\ \label{energy conditins1}
\end{split}
\end{eqnarray}

Using Eqs. (\ref{friedman3}) and (\ref{friedman4}),
 the energy conditions in modified teleparallel gravity model with higher-derivative torsion terms are obtained as
\begin{eqnarray}
\begin{split}
&\textbf{NEC}:\\
&\rho_{eff}+p_{eff}=\rho_m+p_m-\frac{F}{2}-6H^{2}F_{T}-6F_{X_2}\dot{H^{2}}-6HF_{X_2}(3H\dot{H}+\ddot{H})\\
&+30H\dot{H}\dot{F}_{X_2}-432H^{4}F_{X_1}\dot{H}-18H^{3}\dot{F}_{X_2}-2\dot{H}+2\Big[F_{T}\dot{H}
+H\dot{F}_{T}
\\&+24H^{2}\dot{H}\ddot{F}_{X_1}+\dot{H}\ddot{F}_{X_2}H\dddot{F}_{X_2}+24H(2H\ddot{H}+3\dot{H}^{2})\dot{F}_{X_1}+24H^{2}F_{X_1}\dddot{H}
\\
&+96HF_{X_1}\dot{H}\ddot{H}+24F_{X_1}\dot{H^{2}}(12H^{2}+\dot{H})\Big]\geq0,\label{nec}
\end{split}
\end{eqnarray}

\begin{eqnarray}
\begin{split}&\textbf{WEC}: NEC,\\&
\rho_{eff}=\rho_m-\frac{F}{2}+3H^{2}-6H^{2}F_{T}-6F_{X_2}\dot{H^{2}}-6H(3H^{2}-\dot{H})\dot{F}_{X_2}\\
&-144H^{3}\dot{H}\dot{F}_{X_1}-6H^{2}\ddot{F}_{X_2}-6H(F_{X_2}+24H^{2}F_{X_1})(3H\dot{H}+\ddot{H})\geq0, \label{energy conditins}
\end{split}
\end{eqnarray}

\begin{eqnarray}
\begin{split}
&\textbf{SEC}: NEC,\\&
\rho_{eff}+3p_{eff}=\rho_{m}+3p_{m}-\frac{F}{2}-6H^{2}(1+F_{T})-6F_{X_2}\dot{H^{2}}-6HF_{X_2}(3H\dot{H}+\ddot{H})\\&
-432H^{4}F_{X_1}\dot{H}+288H^{3}\ddot{H}F_{X_1}+432H^{3}\dot{H}\dot{F}_{X_1}+78H\dot{H}\dot{F}_{X_2}-18H^{3}\dot{F}_{X_2}\\&
-6H^{2}\ddot{F}_{X_2}-6\dot{H}+6\Big[F_{T}\dot{H}+H\dot{F}_{T}
+24H(2H\ddot{H}+3\dot{H}^{2})\dot{F}_{X_1}+24H^{2}\dot{H}\ddot{F}_{X_1}\\&
+\dot{H}\ddot{F}_{X_2}+24H^{2}F_{X_1}\dddot{H}+H\dddot{F}_{X_2}+24F_{X_1}\dot{H^{2}}(12H^{2}+\dot{H})+96HF_{X_1}\dot{H}\ddot{H}\Big]\geq0,
\end{split}
\end{eqnarray}

\begin{eqnarray}
\begin{split}
&\textbf{DEC}: WEC,\\&
\rho_{eff}-p_{eff}=\rho_{m}-p_{m}-\frac{F}{2}+6H^{2}(1-F_{T})-6F_{X_2}\dot{H^{2}}-6HF_{X_2}(3H\dot{H}+\ddot{H})\\&
-432H^{4}F_{X_1}\dot{H}-288H^{3}F_{X_1}\ddot{H}-288H^{3}\dot{H}\dot{F}_{X_1}-6H(3H^{2}-\dot{H})\dot{F}_{X_2}+2\dot{H}\\&
-2\Big[F_{T}\dot{H}+H\dot{F}_{T}+24H[2H\ddot{H}+3\dot{H}^{2}]\dot{F}_{X_1}-6H^{2}\ddot{F}_{X_2}+24H^{2}\dot{H}\ddot{F}_{X_1}\\&
+\dot{H}\ddot{F}_{X_2}+24H^{2}F_{X_1}\dddot{H}+H\dddot{F}_{X_2}+24F_{X_1}\dot{H^{2}}(12H^{2}+\dot{H})\\&
+96HF_{X_1}\dot{H}\ddot{H}\Big]\geq0\,.\label{dec}
\end{split}
\end{eqnarray}
To get some insight on the meaning of the above energy
conditions, in the next section, we consider two specific functions for the Lagrangian (\ref{action}),
to obtain the constraints on the parametric space of the model.

\section{Constraints on specific Models}
In order to analyze the torsional modified gravity model with higher-derivative terms
from the point of view of energy conditions, we use the standard terminology
in studying energy conditions for modified gravity theories. To this end,
we investigate such energy bounds in terms of the cosmographic parameters \cite{Capozziello08},
i.e. the Hubble, deceleration, jerk, snap and lerk  parameters, defined
respectively as
\begin{equation}
q=-\frac{1}{H^{2}}\frac{a^{(2)}}{a},\qquad\ j=\frac{1}{H^{3}}
\frac{a^{(3)}}{a},\qquad\ s=\frac{1}{H^{4}}\frac{a^{(4)}}{a}, \qquad\ l=\frac{1}{H^{5}}\frac{a^{(5)}}{a},\label{cosmographic}
\end{equation}
where the superscripts represent the derivative with respect to time.
In terms of these parameters, the Hubble parameter as well as its higher time derivatives are given by
\begin{eqnarray}
\begin{split}
&H^{(1)}=-H^{2}(1+q),\\&
H^{(2)}=H^{3}(j+3q+2),\\&
H^{(3)}=H^{4}(s-2j-5q-3),\\&
H^{(4)}=H^{5}[l-5s+10(q+2)j+30(q+2)q+24]\\&\label{hubble-dot}
\end{split}
\end{eqnarray}
respectively. The results of energy conditions in terms of cosmographic parameters for our modified $f(T)$ gravity model
 can be achieved from the constraints (\ref{nec})-(\ref{dec}).

\subsection{\textbf{Model I:} $F(T,X_{1},X_{2})=T+\frac{\alpha_{1}X_{1}}{T^{2}}+\alpha_{2}e^{\frac{\delta X_{1}}{T^{4}}}$}

\begin{figure}[h]
\epsfig{figure=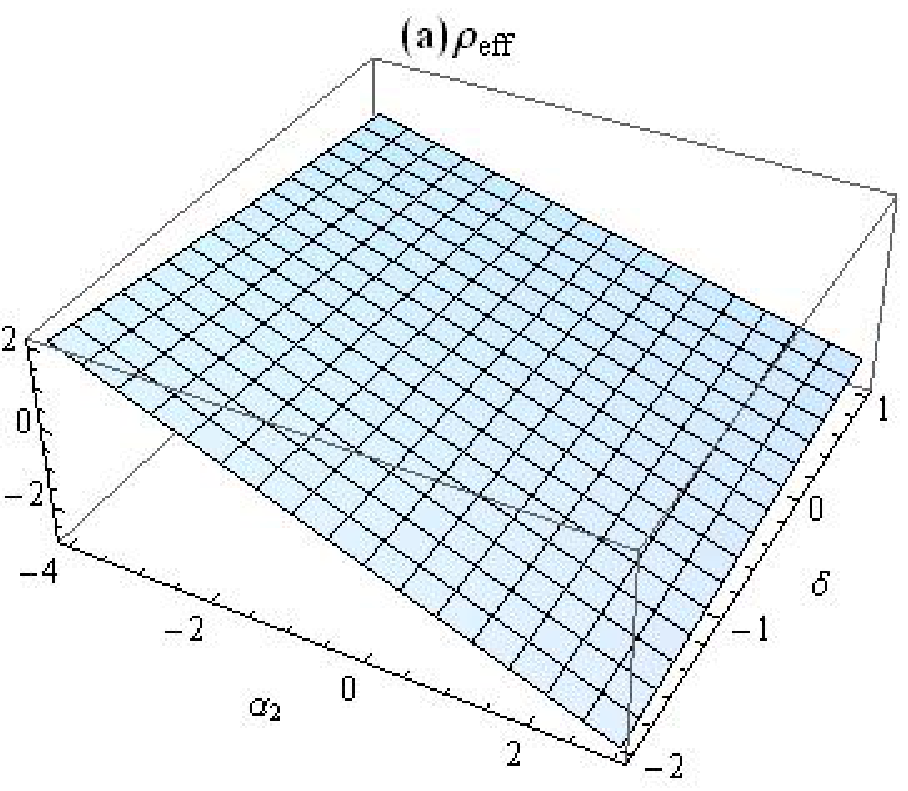,width=6cm}\epsfig{figure=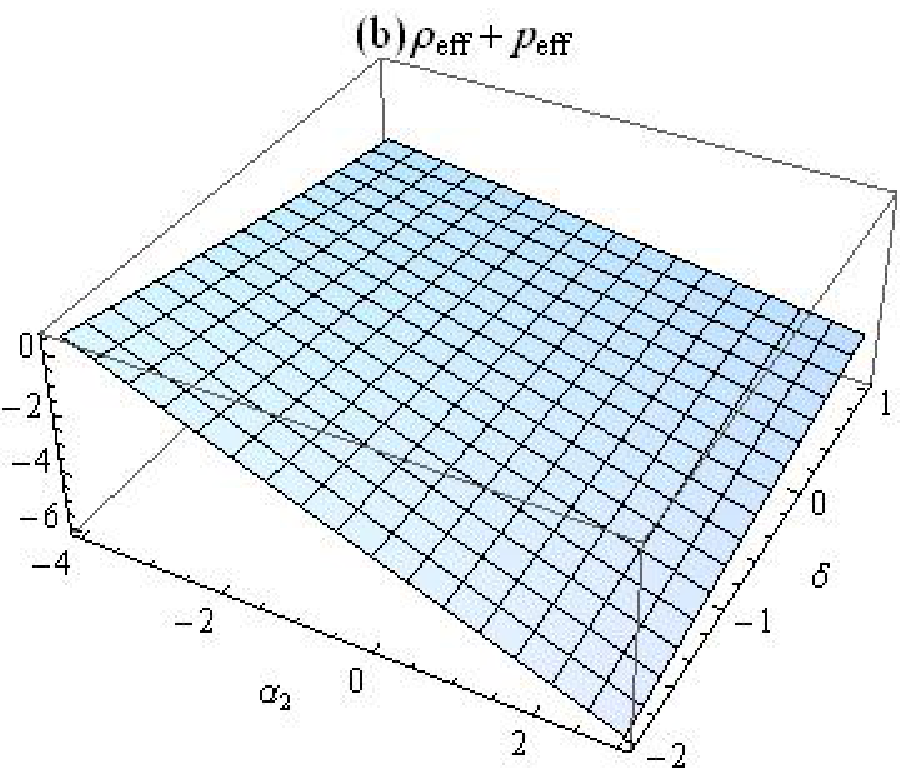,width=6cm}
\vspace*{6pt}\caption{Plot of the weak energy condition for the specific form given by Equ. (\ref{func1}).
Plot (a) represents $\rho_{eff}\geq0$ versus $\alpha_2$ and $\delta$ with $\alpha_1=-2$ . Plot (b) shows
$\rho_{eff}+p_{eff}\geq0$ with the same parameters as plot (a).
 The positivity requirement of the weak energy condition is satisfied in the plots
for the parameter range considered.\protect\label{model1}}
\end{figure}

In this subsection, we adopt a specific function for the modified teleparallel Lagrangian (\ref{action}) as \cite{Otalora16}
\begin{equation}\label{func1}
F(T,X_{1},X_{2})=T+\frac{\alpha_{1}X_{1}}{T^{2}}+\alpha_{2}e^{\frac{\delta X_{1}}{T^{4}}},
\end{equation}
  where $\alpha_{1}$, $\alpha_{2}$ and $\delta$ are constants.
  It has been shown that for a wide range
of the model parameters the universe can result in
a dark-energy dominated, accelerating universe and the model can describe the thermal
history of the universe, i.e. the successive sequence
of radiation, matter and dark energy epochs, which is a
necessary requirement for any realistic scenario.
  Since for a theoretical model to be cosmologically viable, it should satisfy
at least the weak energy condition, we examine specially the weak energy
condition in our analysis. More ever, for simplicity we
consider vacuum, i.e. $\rho_m=p_m=0$. Inserting the cosmographical
parameters (\ref{hubble-dot}) in Equ. (\ref{energy conditins}), the bounds on the
model parameters imposed by the weak energy condition are given by
\begin{eqnarray}
\begin{split}
&\rho_{eff}=-10\alpha_{1}H_0^{2}(1+q_0)^{2}-\frac{\alpha_{2}}{2}e^{\frac{\delta (1+q_0)^{2}}{9H_0^{2}}}\Big[1+\frac{24\delta (1+q_0)^{2}}{27H_0^{2}}\Big]\\&
-9H_0^{2}(j_0-1)\Big[36\alpha_{1}+\alpha_{2}\delta e^{\frac{\delta (1+q_0)^{2}}{9H_0^{2}}}\Big]+16H_{0}^{2}\alpha_{1}(1+q_0)^{2}\\&
+\Big(\frac{\alpha_{2}\delta^2}{11664H_0^{11}}e^{\frac{\delta (1+q_0)^{2}}{9H_0^{2}}}\Big)\Big[-288H_0^{7}(1+q_0)^{2}(j_0+3q_0+2)\\&
-288H_0^{7}(1+q_0)^{4}+\frac{8}{9}\frac{(1+q_0)^{4}}{H_0}+\frac{72H_0^{4}}{\delta}(1+q_0)\Big]\geqslant0,
\end{split}
\end{eqnarray}
and
\begin{eqnarray}
\begin{split}
&\rho_{eff}+p_{eff}=\frac{H_0^{2}}{18}\Big(108\alpha_{1}q_0^{2}-72\alpha_{1}q_0^{3}+408\alpha_{1}q_0+96\alpha_{1}j_0q_0+24\alpha_{1}s_0\\&+408\alpha_{1}-54\Big)
-\frac{1}{4374H_0^{6}}\Big(2187\alpha_{2}H_0^{6}-16\alpha_{2}\delta^{3}j_0+72\alpha_{2}\delta^{3}q_0+216\alpha_{2}\delta^{2}H_0^{2}\\&
+8\alpha_{2}\delta^{3}+72\alpha_{2}\delta^{3}q_0^{7}+360\alpha_{2}\delta^{3}q_0^{6}+768\alpha_{2}\delta^{3}q_0^{5}+912\alpha_{2}\delta^{3}q_0^{4}+972\alpha_{2}\delta H_0^{4}q_0\\&
+656\alpha_{2}\delta^{3}q_0^{3}+8\alpha_{2}\delta^{3}j_0^{2}+288\alpha_{2}\delta^{3}q_0^{2}-972\alpha_{2}\delta H_0^{4}-1296\alpha_{2}\delta H_0^{4}j_0\\&
-1080\alpha_{2}\delta^{2}H_0^{2}j_0q_0^{3}+108\alpha_{2}\delta^{2}H_0^{2}j_0^{2}q_0-2448\alpha_{2}\delta^{2}H_0^{2}j_0q_0^{2}+4536\alpha_{2}\delta^{2}H_0^{2}q_0^{2}\\&
-36\alpha_{2}\delta^{2}H_0^{2}q_0^{2}s_0-1944\alpha_{2}\delta H_0^{4}j_0q_0-1872\alpha_{2}\delta^{2}H_0^{2}j_0q_0+108\alpha_{2}\delta^{2}H_0^{2}j_0^{2}\\&
-72\alpha_{2}\delta^{2}H_0^{2}q_0s_0-162\alpha_{2}\delta H_0^{4}s_0-96\alpha_{2}\delta^{3}j_0q_0-240\alpha_{2}\delta^{3}j_0q_0^{2}\\&
-504\alpha_{2}\delta^{2}H_0^{2}j_0+1512\alpha_{2}\delta^{2}H_0^{2}q_0-36\alpha_{2}\delta^{2}H_0^{2}s_0+24\alpha_{2}\delta^{3}j_0^{2}q_0-304\alpha_{2}\delta^{3}j_0q_0^{3}\\&
-48\alpha_{2}\delta^{3}j_0q_0^{5}+1728\alpha_{2}\delta^{2}H_0^{2}q_0^{5}+8\alpha_{2}\delta^{3}j_0^{2}q_0^{3}+4860\alpha_{2}\delta H_0^{4}q_0^{2}+24\alpha_{2}\delta^{3}j_0^{2}q_0^{2}\\&
-192\alpha_{2}\delta^{3}j_0q_0^{4}+5508\alpha_{2}\delta^{2}H_0^{2}q_0^{4}+3402\alpha_{2}\delta H_0^{4}q_0^{3}+7020\alpha_{2}
\delta^{2}H_0^{2}q_0^{3}\Big)e^{\frac{\delta (1+q_0)^{2}}{9H_0^{2}}}\geqslant0
\end{split}
\end{eqnarray}
respectively. The subscript $0$ stands for the present value of the cosmographic quantities.
 Now, we take the following observed values for the $H_0=0.72$, $q_0=-0.64$, $j_0=1.02$, $s_0=-0.39$ and $l_0=4.05$ \cite{Capozziello11}.
 The numerical results for satisfying the weak energy condition
 are given in figure \ref{model1}. For the parametric spaces of the model, we have fixed the value of $\alpha_1$ to $-2$ and
 plot the $\rho_{eff}$ and $\rho_{eff}+p_{eff}$ versus $\alpha_2$ and $\delta$. As the figure shows, the WEC
 is satisfied in the specific form of Equ. (\ref{func1}) for a suitable choice of the subspaces of the model parametric space.

\subsection{\textbf{Model II:} $F(T,X_{1},X_{2})=T+\frac{\beta_{1}X_{2}}{T}+\frac{\beta_{2}
X_{2}^{2}}{T^{3}}+\beta_{3}e^{\frac{\sigma X_{2}}{T^{3}}}$}
\begin{figure}[h]
\epsfig{figure=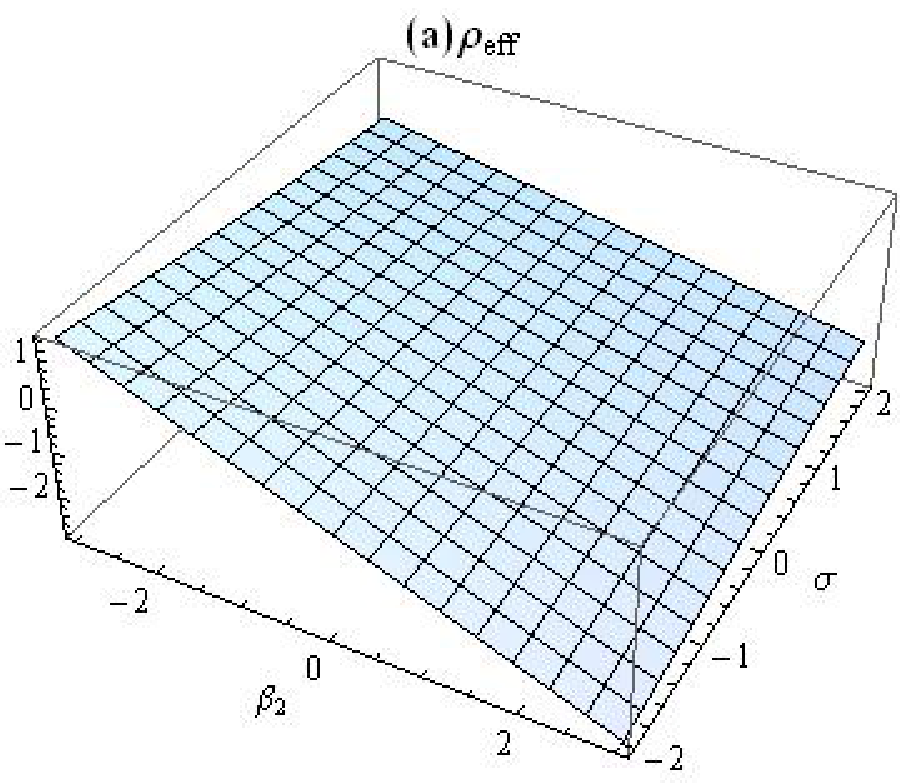,width=6.6cm}\epsfig{figure=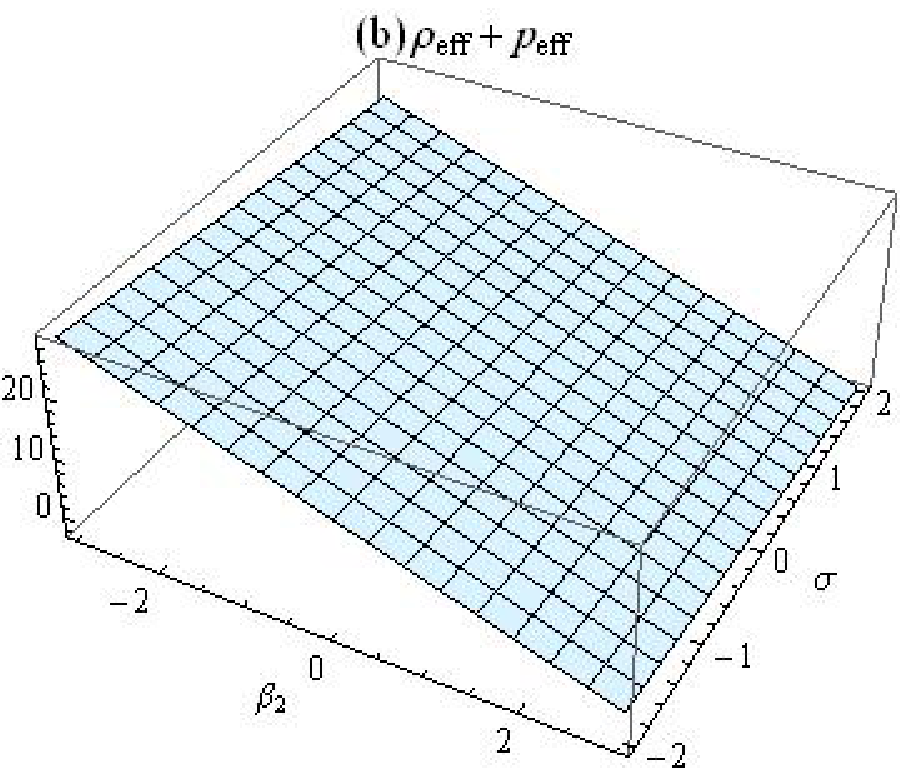,width=6.3cm}
\vspace*{6pt}\caption{The plots depict the weak energy condition versus $\beta_{2}$
and $\sigma$ for the specific form (\ref{func2}). plot (a) corresponds
to $\rho_{eff}\geq0$; Plot (b) is depicted for $\rho_{eff}+p_{eff}\geq0$.
 We have considered the values $\beta_{1}=-4$ and $\beta_{3}=1$.\protect\label{model2}}
\end{figure}

In second case, we consider a class of models in which the action does not depend on $X_{1}
\equiv(\nabla T)^{2}$  but only on $X_{2}\equiv \Box T$ given by the following functional form \cite{Otalora16}
\begin{equation}\label{func2}
F(T,X_{1},X_{2})=T+\frac{\beta_{1}X_{2}}{T}+\frac{\beta_{2}X_{2}^{2}}{T^{3}}+\beta_{3}e^{\frac{\sigma X_{2}}{T^{3}}},
\end{equation}
where $\beta_{1}$, $\beta_{2}$, $\beta_{3}$ and $\sigma$ are constants.
It has been found that in this model the universe will be led to a dark energy
dominated, accelerating phase for a wide region of the parameter space. The
 scale factor behaves asymptotically either as a power law or as an exponential law,
 while for large parameter regions the exact value of the dark-energy equation-of-state
parameter can be in great agreement with observations \cite{Otalora16}.
To examine the model via energy conditions, in a similar procedure to the previous subsection, we consider the vacuum.
Using the cosmographical parameters as before, the condition to justification of the WEC are obtained as
\begin{eqnarray}
\begin{split}
&\rho_{eff}=-\frac{\beta_{3}}{54H_0^{2}}e^{\frac{\sigma (q_0^{2}+j_0+2q_0)}{18H_0^{2}}}\Big(27 H_0^{2}+12\sigma^{2}q_0^{4}+36\sigma^{2}q_0^{3}
+36\sigma^{2}q_0^{2}+4\sigma^{2}j_0\\&+12\sigma^{2}q_0
+2\sigma^{2}s_0+252\sigma H_0^{2}+9\sigma q_0^{2}+9\sigma j_0+18\sigma q_0+6\sigma^{2}j_0q_0^{2}\\&
+216\sigma H_0^{2}q_0^{2}+10\sigma^{2}j_0q_0+2\sigma^{2}q_0s_0-36\sigma H_0^{2}j_0+432\sigma H_0^{2}q_0\Big)\\&-\frac{1}{3}\Big(-288\beta_{2}H_0^{4}q_0^{4}
+144\beta_{1}H_0^{4}q_0^{2}
+216\beta_{1}H_0^{4}+288\beta_{1}H_0^{4}q_0-912H_0^{4}q_0^{2}\\&
-144\beta_{2}H_0^{4}j_0-384\beta_{2}H_0^{4}q_0
-48\beta_{2}H_0^{4}s_0-864\beta_{2}H_0^{4}q_0^{3}+48\beta_{2}H_0^{4}j_0^{2}\\&
+9\beta_{1}H_0^{2}q_0^{2}+9\beta_{1}H_0^{2}j_0+18\beta_{1}H_0^{2}q_0-7\beta_{2}H_0^{2}q_0^{4}
-28\beta_{2}H_0^{2}q_0^{3}-7\beta_{2}H_0^{2}j_0^{2}\\&-28\beta_{2}H_0^{2}q_0^{2}-96\beta_{2}H_0^{2}j_0q_0^{2}
-48\beta_{2}H_0^{4}q_0s_0-72\beta_{1}H_0^{4}j_0\\&
-14\beta_{2}H_0^{2}j_0q_0^{2}-28\beta_{2}H_0^{2}j_0q_0-144\beta_{2}H_0^{4}j_0q_0\Big)\geqslant0\,,
\end{split}\label{rho2}
\end{eqnarray}

\begin{eqnarray}
\begin{split}
&\rho_{eff}+p_{eff}=\frac{\beta_{3}}{1458H_0^{4}}e^{\frac{\sigma(q_0^{2}+j_0+2q_0)}{18H_0^{2}}}\Big(+36\sigma^{3}q_0^{7}
+180\sigma^{3}q_0^{6}+414\sigma^{2}H_0^{2}j_0q_0-729H_0^{4}\\&
+360\sigma^{3}q_0^{5}+360\sigma^{3}q_0^{4}+180\sigma^{3}q_0^{3}+4\sigma^{3}j_0^{2}+36\sigma^{3}q_0^{2}+\sigma^{3}s_0^{2}+10\sigma^{3}j_0q_0s_0\\&
-5832\sigma H_0^{4}-486\sigma H_0^{2}q_0^{2}-324\sigma H_0^{2}j_0-648\sigma q_0H_0^{2}+36\sigma^{3}j_0q_0^{5}-4536\sigma H_0^{4}j_0q_0\\&
+1728\sigma^{2}H_0^{2}q_0^{5}+9\sigma^{3}j_0^{2}q_0^{3}+132\sigma^{3}j_0q_0^{4}+12\sigma^{3}q_0^{4}s_0+5508\sigma^{2}H_0^{2}q_0^{4}+8100\sigma H_0^{4}q_0^{3}\\&
+21\sigma^{3}j_0^{2}q_0^{2}+180\sigma^{3}j_0q_0^{3}+36\sigma^{3}q_0^{3}s_0+6588\sigma^{2}H_0^{2}q_0^{3}+7776\sigma H_0^{4}q_0^{2}+16\sigma^{3}j_0^{2}q_0\\&
+108\sigma^{3}j_0q_0^{2}+36\sigma^{3}q_0^{2}s_0+\sigma^{3}q_0s_0^{2}-126\sigma^{2}H_0^{2}j_0^{2}+3564\sigma^{2}H_0^{2}q_0^{2}-3240\sigma H_0^{4}j_0\\&
-5184\sigma H_0^{4}q_0-324\sigma H_0^{4}s_0-81\sigma H_0^{2}q_0^{3}+24\sigma^{3}j_0q_0+4\sigma^{3}j_0s_0+81\sigma H_0j_0\\&
+12\sigma^{3}q_0s_0+180\sigma^{2}H_0^{2}j_0+18\sigma^{2}H_0^{2}l_0+756\sigma^{2}H_0^{2}q_0+198\sigma^{2}H_0^{2}s_0+81\sigma H_0q_0^{2}\\&
+162\sigma H_0q_0+270\sigma^{2}H_0^{2}j_0q_0^{3}+6\sigma^{3}j_0q_0^{2}s_0
-162\sigma^{2}H_0^{2}j_0^{2}q_0+468\sigma^{2}H_0^{2}j_0q_0^{2}\\&
-36\sigma^{2}H_0^{2}j_0s_0+18\sigma^{2}H_0^{2}l_0q_0+468\sigma^{2}H_0^{2}q_0s_0
-81\sigma H_0^{2}j_0q_0+306\sigma^{2}H_0^{2}q_0^{2}s_0\Big)\\&
+\frac{H_{0}}{18}\Big(-23328\beta_{1}H_0^{3}j_0q_0+288\beta_{2}H_0^{3}j_0s_0-384\beta_{2}H_0^{3}q_0^{2}s_0+216\beta_{2}H_0j_0q_0\\&
+102\beta_{2}H_0q_0^{4}-90\beta_{1}H_0q_0^{2}-132\beta_{1}H_0q_0+216\beta_{2}H_0q_0^{2}+264\beta_{2}H_0q_0^{3}\\&
-66\beta_{1}H_0j_0+156\beta_{2}H_0j_0q_0^{2}-18H_0+288\beta_{1}H_0^{3}j_0\\&
+4608\beta_{2}H_0^{3}j_0q_0^{2}+24\beta_{2}H_0j_0q_0^{3}+12\beta_{2}H_0j_0^{2}q_0-12\beta_{1}H_0j_0q_0-24\beta_{2}j_0q_0^{2}\\&
+1536\beta_{2}H_0^{3}j_0q_0^{3}+480\beta_{2}H_0^{3}j_0^{2}q_0+12\beta_{2}H_0q_0^{5}-12\beta_{1}H_0q_0^{3}-12\beta_{2}q_0^{4}\\&
+12\beta_{1}q_0^{2}-12\beta_{2}j_0^{2}-48\beta_{2}q_0^{2}+12\beta_{1}j_0+24\beta_{1}q_0-48\beta_{2}j_0q_0\\&
+144\beta_{1}H_0^{3}q_0^{3}-144\beta_{1}H_0^{3}s_0-2592\beta_{1}H_0^{3}-1728\beta_{1}H_0^{3}q_0^{2}-4032\beta_{1}H_0^{3}q_0\\&
+4992\beta_{2}H_0^{3}q_0^{3}-96\beta_{2}H_0^{3}j_0^{2}+8832\beta_{2}H_0^{3}q_0^{2}+1920\beta_{2}H_0^{3}j_0-96\beta_{2}H_0^{3}l_0\\&
+4032\beta_{2}H_0^{3}q_0+96\beta_{2}H_0^{3}s_0-2496\beta_{2}H_0^{3}q_0^{4}-48\beta_{2}q_0^{3}-2400\beta_{2}H_0^{3}q_0^{5}\\&
-96\beta_{2}H_0^{3}l_0q_0+54\beta_{2}H_0j_0^{2}+4128\beta_{2}H_0^{3}j_0q_0\Big)\geqslant0\,.
\end{split}\label{rho-p2}
\end{eqnarray}
From Eqs. (\ref{rho2}) and (\ref{rho-p2}) it is clear that the WEC is dependent
on a wide range of parameters, i.e. $\beta_{1}$, $\beta_{2}$,
$\beta_{3}$ and $\sigma$. So to analyse the energy conditions, we consider specific values for some of the parameters.
In particular, we take $\beta_{1}=-4$ and $\beta_{3}=1$ and plot the
WEC as a function of $\beta_{2}$ and $\sigma$. Figure \ref{model2} shows the behavior of the $\rho_{eff}$ and
$\rho_{eff}+p_{eff}$ versus the free parameters $\beta_{2}$ and $\sigma$. As the figure shows, the specific model II
 is consistent with the WEC inequalities in the subspaces of the model parametric space.
\section{Conclusion}
The modified teleparallel gravity with higher-derivative torsion terms is a generalization of $f(T)$ gravity
theories constructed from the higher-derivative, $(\nabla T)^{2}$ and $\Box T$ terms,
i.e. theories that are characterized by the Lagrangian $F(T,(\nabla T)^{2},\Box T)$.
It has been shown that this model can result in
an effective dark energy sector that comprises of the novel torsional
contributions due to the higher-order derivative terms.

In this paper, we have developed some constraints on this generalized $f(T)$ model by
examining the respective energy conditions. In this respect, we have written the WEC, DEC, NEC and SEC
 in terms of an effective energy-momentum tensor which arises from the further degrees of freedom
 related to the higher-derivative terms. In order to illustrate how these conditions can constrain the
model, we have considered two specific functional forms of the Lagrangian
that account for accelerating phase of the universe for a wide range of their parameter spaces.

Specially, we have examined the validity of WEC to obtain bounds on the free parameters of both models. Rephrasing the WEC
inequalities in terms of the present-day values of the cosmographic parameters,
we have found the constraints on the free parameter of the presented models. As a result, the WEC is fulfilled in both models
 in some subspaces of the model parametric space.
\section{Acknowledgment}
The authors would like to thank professor Kourosh Nozari for useful comments and discussions.


\begin{thebibliography}{}

\bibitem{Riess98} A. G. Riess {\it et al.},   Astron.\ J.\  {\bf 116},
1009 (1998)

\bibitem{Perlmutter99}
  S. Perlmutter {\it et al.},  Astrophys.\ J.\  {\bf 517}, 565
(1999)

\bibitem{Ade15}
P. A. R. Ade et al. [Planck Collaboration], Planck 2015;
arXiv:1502.01589 [astro-ph.CO].

\bibitem{Peebles03}
P.J.E, Peebles, B. Ratra,  Rev. Mod. Phys. {\bf75}, 559 (2003)

\bibitem{Li02}
X.Z. Li, , J.G. Hao, ,D.J. Liu,  class. Quantum Grav. {\bf19}, 6049 (2002)

\bibitem{Caldwell02}
R.R. Caldwell, Phys. Lett. B {\bf545}, 23
(2002)

\bibitem{Kamenshchik01}
A. Y. Kamenshchik, U. Moschella and V. Pasquier, Phys. Lett.
B {\bf511} 265 (2001); arXiv:gr-qc/0103004.
\bibitem{Copeland06}
E. J. Copeland, M. Sami and S. Tsujikawa, Int. J. Mod. Phys. D {\bf15}, 1753 (2006).
\bibitem{Sotiriou10}
T. P. Sotiriou and V. Faraoni, Rev. Mod. Phys., {\bf82}, 451 (2010).
\bibitem{Nojiri11}
S. Nojiri and S. D. Odintsov, Phys. Rept. {\bf505}, 59 (2011).
\bibitem{Clifton12}
T. Clifton, P.G. Ferreira, A. Padilla and C. Skordis, Phys.Rept {\bf513}, 1 (2012)

\bibitem{Einstein28}
A. Einstein and S.P.A. Wiss. phys. {\bf217}, 224 (1928)

\bibitem{Hayashi79}
K. Hayashi and T. Shirafuji, Phys.Rev. D {\bf19}, 3524 (1979)
\bibitem{Ferraro07}
R. Ferraro and F. Fiorini, Phys. Rev. D  {\bf75}, 084031 (2007)

\bibitem{Bengochea09}
G. R. Bengochea and R. Ferraro, Phys. Rev. D {\bf79}, 124019 (2009)
\bibitem{Linder10}
E. V. Linder, Phys. Rev. D {\bf81}, 127301 (2010)
\bibitem{Cai16}
Y.F Cai, S. Capozziello, M.D. Laurentis and E.N. Saridakis, Rept. Prog. Phys., {\bf79} 106901 (2016)
\bibitem{Otalora16}
G. Otalora and E. N. Saridakis, Phys. Rev. D {\bf94}, 084021 (2016)
\bibitem{Dadhich07}
N. Dadhich, [arXiv:gr-qc/0511123]
\bibitem{Kar07}
S. Kar and S. Sen Gupta, Pramana {\bf69}, 49 (2007)
\bibitem{Ber06}
 S. E. Bergliaffa, Phys. Lett. B {\bf642}, 311 (2006)
\bibitem{Sant07}
 J. Santos, J. S. Alcaniz, M. J. Reboucas and F. C. Carvalho, Phys. Rev. D {\bf76}, 083513 (2007)
\bibitem{Ata09}
 K. Atazadeh, A. Khaleghi, H. R. Sepangi and Y. Tavakoli, Int. J. Mod. Phys. D {\bf18}, 1101 (2009)
\bibitem{Capozziello15}
S. Capozziello , F.S. N. Lobo and J. P. Mimoso, Phys.Rev. D{\bf91} 124019 (2015)
\bibitem{Bertolami09}
O. Bertolami and M. C. Sequeira, Phys. Rev. D {\bf79}, 104010 (2009)
\bibitem{Garcia10}
N. M. Garcia and F. S. N. Lobo, Phys. Rev. D {\bf82}, 104018 (2010)
\bibitem{Wang10}
J. Wang, Y. B. Wu, Y. X. Guo, W. Q. Yang and L. Wang, Phys. Lett. B {\bf689}, 133 (2010)
\bibitem{Garcia11}
 N. M. Garcia and F. S. N. Lobo, Class. Quant. Grav. {\bf28}, 085018 (2011)
\bibitem{Wang12}
 J. Wang and K. Liao, Class. Quant. Grav. {\bf29}, 215016 (2012)
\bibitem{Alvarenga13}
 F. G. Alvarenga, M. J. S. Houndjo, A. V. Monwanou and J. B. C. Orou, Int. J. Mod. Phys. {\bf4}, 130
(2013)
\bibitem{Haghani13}
Z. Haghani, T. Harko, F. S. N. Lobo, H. R. Sepangi and S. Shahidi, Phys. Rev. D {\bf88}, 044023(2013)
\bibitem{Odintsov13}
S. D. Odintsov and D. Saez-Gomez, Phys. Lett. B {\bf725}, 437 (2013)
\bibitem{Sharif13}
 M. Sharif and M. Zubair, Phys. Soc. Jap. {\bf82}, 014002 (2013)
 \bibitem{Wu10}
 P. Wu and H. Yu, Mod. Phys. Lett. A {\bf25}, 2325 (2010).
\bibitem{Garcia11a}
 N. M. Garcia, T. Harko, F. S. N. Lobo and J. P. Mimoso, Phys. Rev. D {\bf83}, 104032 (2011)
\bibitem{Garcia11b}
 N. Montelongo Garcia, F. S. N. Lobo, J. P. Mimoso and T. Harko, Phys. Conf. Ser. {\bf314}
012056 (2011)
\bibitem{Bani12}
A. Banijamali, B. Fazlpour and M. R. Setare, Ast. phys. Space Sci. {\bf338} 327 (2012)
\bibitem{Sharif13a}
Sharif, M. and Waheed, S. Advances in High Energy Phys., {\bf2013} 253985 (2013)
\bibitem{Capozziello14}
S. Capozziello, Francisco S. N. Lobo and J. P. Mimoso, Phys. Lett. B {\bf730} 280 (2014)
 \bibitem{Liu12}
 D. Liu and M. J. Reboucas, Phys. Rev. D {\bf86}, 083515 (2012)
\bibitem{Boehmer12}
 C. G. Boehmer, T. Harko and F. S. N. Lobo, Phys. Rev. D {\bf85}, 044033 (2012)
\bibitem{Jamil13}
 M. Jamil, D. Momeni and R. Myrzakulov, Gen. Rel. Grav. {\bf45}, 263 (2013)
\bibitem{Kiani14}
F. Kiani, Kourosh Nozari, Phys. Lett. B {\bf728}, 554 (2014)
\bibitem{Zubair15}
M. Zubair, Saira Waheed, Ast. phys. Space Sci. {\bf355}, 361 (2015)
\bibitem{Jawad15}
A.Jawad, Phys. J. Plus {\bf94}, 130 (2015)
\bibitem{Sharif16}
M. Sharif, Ayesha Ikram, Eur. Phys. J. C {\bf76}, 640 (2016)
\bibitem{Capozziello08}
S. Capozziello, V.F. Cardone and V. Salzano, Phys. Rev. D {\bf78}, 063504 (2008)
\bibitem{Capozziello11}
S. Capozziello, V. F. Cardone, H. Farajollahi and A. Ravanpak, Phys. Rev. D {\bf84}, 043527 (2011)
\end{thebibliography}
\end{document}